\documentclass[twocolumn,showpacs,preprintnumbers,amsmath,amssymb]{revtex4}

\usepackage{graphicx}
\usepackage{dcolumn}
\usepackage{bm}

\begin{document}
\title{Dynamical Coulomb blockade and spin-entangled electrons}
\author{Patrik Recher and Daniel Loss}
\address{ Department of Physics and Astronomy, University of
Basel,\\
Klingelbergstrasse 82, CH-4056 Basel, Switzerland}
 
\date{\today}

 
\begin{abstract}
We consider the production of mobile and nonlocal pairwise spin-entangled electrons from tunneling of a BCS-superconductor (SC) to two normal Fermi liquid leads. The necessary mechanism to separate the two electrons coming from the same Cooper pair (spin-singlet) is achieved by coupling the SC to leads with a finite resistance. The resulting dynamical Coulomb blockade effect, which we describe phenomenologically in terms of an electromagnetic environment, is shown to be enhanced for tunneling of two spin-entangled electrons into the same lead compared to the process where the pair splits and each electron tunnels into a different lead. On the other hand in the pair-split process, the spatial correlation of a Cooper pair leads to a current suppression as a function of distance between the two tunnel junctions which is weaker for effectively lower dimensional SCs.
\end{abstract}
\narrowtext
\maketitle
{\it Introduction.--\/}
The controled creation of nonlocal entanglement is crucial in quantum communication as well as in quantum computation tasks \cite{Chuang}.
Quantum entanglement is further interesting in its own right since it leads to a  violation of Bell's inequality \cite{Bell}. Several solid state entanglers, a device that creates mobile  and nonlocal pairwise entangled electrons, were proposed recently \cite{RSL,RL,Yamamoto,tanglers,Bena,Saraga}. A particularly interesting quantity  is the spin of the electron which was shown to be a promising realization of a quantum bit \cite{Loss}. A natural source of spin entanglement is provided by Cooper pairs in an s-wave superconductor (SC), since the Cooper pairs are in a spin singlet state. Weakly coupling the SC to a normal region allows for (pair)-tunneling of Cooper pairs from the SC to normal leads and single particle tunneling is suppressed at low energies below the SC gap. Subsequently, Coulomb interaction between the two electrons of a pair can be used to separate them spatially leading to nonlocality \cite{Choi}. To mediate the necessary interaction entangler setups containing quantum dots \cite{RSL} or which exhibit Luttinger liquid correlations \cite{RL,Bena} (e.g. nanotubes in the metallic regime) were proposed recently. \\
In this letter we show that a considerably simpler experimental realization can be used to generate the necessary Coulomb interaction between the electrons of a pair. Indeed  if the normal leads are resistive a dynamical Coulomb blockade (CB) effect is generated with the consequence that in a pair tunneling process into the same lead the second electron still experiences the Coulomb repulsion of the first one, which has not yet diffused away. Natural existing candidates with long spin decoherence lengths ($\sim 100$ $\mu$m \cite{Awschalom})  for such a setup are e.g. semiconductor systems tunnel-coupled to a SC, as experimentally implemented in InAs \cite{InAs}, InGaAs \cite{Franceschi} or GaAs/AlGaAs \cite{Marsh}. Recently, 2DEGs with a resistance per square up to almost the quantum resistance $R_{Q}=h/e^{2}\sim 25.8$ k$\Omega$ could be achieved by depleting the 2DEG with a voltage applied between a back gate and the 2DEG \cite{Rimberg}. In metallic normal NrCr leads of width $\sim 100$ nm and length $\sim 10$ $\mu$m, resistances of $R=22-24$ k$\Omega$ have been produced at low temperatures. Even larger resistances $R=200-250$ k$\Omega$ have been measured in Cr leads \cite{Kuzmin}.\\
We use  a phenomenological approach to describe charge dynamics in the electromagnetic circuit which is described in terms of normal-lead impedances and junction capacitances, see Fig. 1.
The subgap transport of a single SN- junction under the influence of an electromagnetic environment has been studied in detail \cite{Diener,Huck}. In order to create nonlocal entangled states in the leads we have to go beyond previous work to investigate the physics of {\it two} tunnel junctions in parallel with two distinct transport channels for singlets. A Cooper pair can tunnel as a whole into one lead, or the pair can split and the two electrons tunnel to separate leads, leading to a nonlocal entangled spin-pair in the leads \cite{splitter}. In the case where the pair splits we find that the CB effect provided by the electromagnetic environment is uncorrelated for the two electron charges. In contrast, if the two electrons tunnel into the same lead we find a dynamical CB  consistent with a charge $q=2e$, where $e$ ist the elementary charge.  Thus  the CB effect is twice  as large for the unsplit process which enhances the probability for a nonlocal (pair-split) process. On the other hand we show that the spatial correlations of a Cooper pair results in a suppression factor for tunneling via different junctions which is weaker for lower dimensional SCs.\\
{\it Setup and formalism.--\/}
The setup is sketched in Fig. 1. The SC is held at the (electro-)chemical potential $\mu_{S}$  by a voltage source V. The two electrons of a Cooper pair can tunnel via two junctions placed at points ${\bm r_{1}}$ and ${\bm r_{2}}$ on the SC to two separate normal leads 1 and 2 which have resistances $R_{1}$ and $R_{2}$, {\it resp.} They are kept at the same chemical potential $\mu_{l}$ so that a bias voltage $\mu\equiv \mu_{S}-\mu_{l}$ is applied between SC and leads \cite{leads}.
\begin{figure}[h]
\centerline{\includegraphics[width=5.2cm]{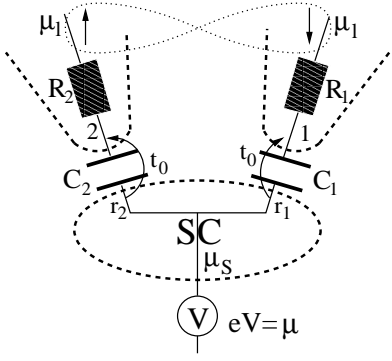}}
\label{fig2} \caption{Entangler setup: A superconductor (SC) with chemical potential $\mu_{S}$ is tunnel-coupled (amplitude $t_{0}$)  via two points ${\bf r}_{1}$ and ${\bf r}_{2}$ of the SC to two Fermi liquid leads 1,2 with resistance $R_{1,2}$. The two leads are held at the same chemical potential $\mu_{l}$ such that a  bias voltage $\mu=\mu_{S}-\mu_{l}$ is applied between the SC and the two leads. The tunnel-junctions 1,2 have capacitances $C_{1,2}$.} 
\end{figure}
The system Hamiltonian decomposes into three parts $
H=H^{e}+H_{env}+H_{T}$. Here $H^{e}=H_{S}+\sum_{n=1,2}H_{ln}$ describes the
electronic parts of the isolated subsystems consisting of the SC and Fermi liquid leads 
$n=1,2$, with $H_{ln}=
\sum_{\bf{p}\sigma}\varepsilon_{p}\,c_{n\bf{p}\sigma}^{\dagger}c_{n\bf{p}\sigma}$,
where
$\sigma=(\uparrow,\downarrow)$. The s-wave SC is described by the BCS-Hamiltonian $H_{S}-\mu_{S}N_{S}=\sum_{\bf{k}\sigma}E_{k}\,\gamma_{\bf{k}\sigma}^{\dagger}\gamma_{\bf{k}\sigma}$ with the  quasiparticle spectrum
$E_{k}=(\xi_{k}^{2}+\Delta^{2})^{1/2}$ where
$\xi_{k}=k^{2}/2m-\mu_{S}$.
The electron creation $(c_{{\bf k}\sigma}^{\dagger})$ and annihilation $(c_{{\bf k}\sigma})$ operators are related to
the quasiparticle operators by  the Bogoliubov transformation
$c_{\bf{k}(\uparrow/\downarrow)}=u_{k}\gamma_{\bf{k}(\uparrow/\downarrow)}\pm v_{k}\gamma_{-\bf{k}(\downarrow/\uparrow)}^{\dagger}$,
where $u_{k}$ and $v_{k}$ are the usual BCS coherence
factors. To describe resistance and dissipation in the normal leads we use a phenomenological approach \cite{Devoret}, where the electromagnetic fluctuations  in the circuit (being bosonic excitations) due to electron-electron interaction and the lead resistances are modeled by a bath of harmonic oscillators which is linearly coupled to the charge fluctuation ${Q_{n}}$ of the junction capacitor $n$ (induced by the tunneling electron). This physics is described by \cite{Leggett,oscillators}
\begin{equation}
\label{env}
 H_{env, n}=
\frac{{ Q_{n}}^{2}}{2C_{n}}+\sum\limits_{j=1}^{N}\,\left[\frac{q_{nj}^{2}}{2C_{nj}}+\frac{({\phi_{n}}-\varphi_{nj})^{2}}{2e^{2}L_{nj}}\right].
\end{equation}
The phase ${\phi_{n}}$ of junction $n$ is the conjugate variable to the charge satisfying $[{\phi_{n}},{Q_{m}}]=ie\delta_{n,m}$.
As a consequence $e^{-i{\phi_{n}}}$ reduces 
${Q_{n}}$ by one elemantary charge $e$. We remark that in
our setup the SC is held at the constant chemical potential
$\mu_{S}$ by the voltage source, see Fig. 1. 
Therefore the charge relaxation of a
non-equilibrium charge on one of the capacitors described by
(\ref{env}) does not influence the charge dynamics of the other
junction and, as a consequence, $H_{env}=\sum_{n=1,2}H_{env,n}$ \cite{comment1}. Electron  tunneling through junctions 1, 2 located  at
points ${\bf r}_{1}$,
 ${\bf r}_{2}$ of the SC nearest to the leads 1,2 is described by the tunneling Hamiltonian
 $H_{T}=\sum_{n=1,2}H_{Tn}+{\rm h.c.}$ where
\begin{equation}
\label{Htunnel}
H_{Tn}=t_{0}\sum\limits_{\sigma}\,\psi_{n\sigma}^{\dagger}\Psi_{\sigma}({\bf
r}_{n})\,e^{-i{\phi}_{n}}.
\end{equation}
Here $t_{0}$ is the bare electron tunneling amplitude which we assume to be spin-independent and the same for both leads. Since $H_{T}$ conserves spin we have $[H,{\bf S}_{tot}^{2}]=0$, and thus the two electrons from a given Cooper pair singlet which have tunneled to the lead(s) remain  in the singlet state.\\
{\it Current of two electrons tunneling into different leads.--\/} 
We use a T-matrix approach \cite{Merzbacher} to calculate tunneling currents. At zero temperature the current $I_{1}$ for tunneling of two
electrons coming from the same Cooper pair into {\it different} leads is given to lowest order in $t_{0}$ by \cite{RL}
\begin{eqnarray}
\label{current1}
 I_{1}&=&2e\sum\limits_{{\tiny
\begin{array}{l}n\not=n'\\m\not=m'\end{array}}}
\int\limits_{-\infty}^{\infty}dt\,\int\limits_{0}^{\infty}dt'\,
\int\limits_{0}^{\infty}dt''\,e^{-\eta(t'+t'')+i(2t-t'-t'')\mu}\nonumber\\
&&\times\langle H_{Tm}^{\dagger}(t-t'')H_{Tm'}^{\dagger}(t)
H_{Tn}(t')H_{Tn'}(0)\rangle,
\end{eqnarray}
where $\eta\rightarrow 0^{+}$, and the expectation value is to be taken in the groundstate of the unperturbed system. The physical interpretation of Eq. (\ref{current1}) is a hopping process of two electrons with opposite spins from two spatial points ${\bf r}_{1}$ and ${\bf r}_{2}$ of the SC to the two leads 1,2, thereby
removing  a Cooper pair in the SC, and back again. The delay times
between the two tunneling processes of the electrons within a pair is
given by $t'$ and $t''$, {\it resp.}, whereas the time between destroying and creating a Cooper pair is given by $t$. This process  is contained in the
correlation function 
\begin{widetext}
\begin{eqnarray}
\label{corr1}
\sum\limits_{{\tiny
\begin{array}{c}n\not=n'\\m\not=m'\end{array}}}\,\langle
H_{Tm}^{\dagger}(t-t'')\,H_{Tm'}^{\dagger}(t)
\,H_{Tn}(t')\,H_{Tn'}(0)\rangle\qquad\qquad\qquad\qquad\qquad&&\nonumber\\
=|t_{0}|^{4}\sum\limits_{\sigma,\,n\not=m}\left\{
{G}_{n\sigma}(t-t'')\,{G}_{m,-\sigma}(t-t')\,{\cal F}_{nm\sigma}(t')\,{\cal F}_{nm\sigma}^{*}(t'')\,
\langle e^{i{\phi_{n}}(t-t'')}\,e^{-i{\phi_{n}}(0)}\rangle\langle e^{i{\phi_{m}}(t-t')}\,e^{-i{\phi_{m}}(0)}\rangle\right.&&\nonumber\\
\left.-\,{G}_{m,-\sigma}(t-t'-t'')\,{G}_{n\sigma}(t)\,{\cal F}_{nm\sigma}(t')\,{\cal F}_{mn,-\sigma}^{*}(t'')\,
\langle e^{i{\phi_{m}}(t-t'-t'')}\,e^{-i{\phi_{m}}(0)}\rangle\langle e^{i{\phi_{n}}(t)}\,e^{-i{\phi_{n}}(0)}\rangle\right\}.&&
\end{eqnarray}
\end{widetext}
The lead Green's functions are ${G}_{n\sigma}(t)\equiv\langle\psi_{n\sigma}(t)\psi_{n\sigma}^{\dagger}(0)\rangle\simeq(\nu_{l}/2)/it$, with $\nu_{l}$ being the DOS per volume at the Fermi level $\mu_{l}$ of the leads. The anomalous Green's function of the SC is ${\cal F}_{{n}{m}\sigma}(t)\equiv\langle\Psi_{-\sigma}({\bf r}_{m},t)\Psi_{\sigma}({\bf r}_{n},0)\rangle=({\rm sgn(\sigma)}/V_{S})\sum_{{\bf k}}u_{k}v_{k}\exp(-iE_{k}t+i{\bf k}\cdot\delta{\bf r})$ with $\delta{\bf r}={\bf r}_{1}-{\bf r}_{2}$, and $V_{S}$ is the volume of the SC.
The bath correlator  can be expressed as $\langle\exp(i{\phi_{n}}(t))\exp(-i{\phi_{n}(0)})\rangle=\exp[J(t)]$ with $J(t)=2\int_{0}^{\infty}(d\omega/\omega) ({\rm Re} Z_{T}(\omega)/R_{Q})(\exp(-i\omega t)-1)$. Here we introduced the total impedance $Z_{T}=(i\omega C+R^{-1})^{-1}$, with a purely Ohmic lead impedance $Z_{n}(\omega)=R$, which we assume to be the same for both tunnel-junctions and leads. For small times, $\omega_{R}|t|\ll 1$, we can approximate $J(t)\sim -iE_{c}t$ where $E_{c}=e^{2}/2C$ is the charging energy and $\omega_{R}=1/RC$ is the bath frequency cut-off which is the inverse classical charge relaxation time $\tau_{cl}$ of an RC-circuit. For the long-time behavior, $\omega_{R}|t|\gg 1$, we get $J(t)\sim -(2/g)[\ln(i\omega_{R}t)+\gamma]$ with $\gamma=0.5772$ the Euler number and $g=R_{Q}/R$ is the dimensionless lead conductance which determines the power-law decay of the bath correlator at long times. \\
We first consider the low bias regime $\mu\ll\Delta,\omega_{R}$.
In this limit the delay times $t'$ and $t''\stackrel{<}{\sim} 1/\Delta$ can be neglected compared to $t\stackrel{<}{\sim}1/\mu$ in all correlators in (\ref{corr1}) and the bath correlators are dominated by the long-time behavior of $J(t)$. We then obtain for the current 
\begin{equation}
\label{I1}
I_{1}=e\pi\mu\Gamma^{2}F_{d}^{2}(\delta r)\frac{e^{-4\gamma/g}}{\Gamma(4/g+2)}\left(\frac{2\mu}{\omega_{R}}\right)^{4/g}.
\end{equation}
The geometrical factor coming from the spatial correlation of a Cooper pair is $F_{d=3}(\delta r)=[\sin(k_{F}\delta r)/k_{F}\delta r]\exp(-\delta r/\pi\xi)$ with $\delta r=|\delta{\bf r}|$. The exponential decay of the correlation sets in on the length scale of the coherence length $\xi$.  It is on the order of micrometers for usual s-wave materials and can be assumed to be larger than $\delta r$ which could reach nanometers. More severe is the power-law decay $\propto 1/(k_{F}\delta r)^{2}$ with $k_{F}$ the Fermi wavenumber in the SC. This power-law is sensitive to the effective dimensionality $d$ \cite{dimension} of the SC with weaker decay in lower dimensions. Indeed, in  two dimensions and for $k_{F}\delta r \gg 1$, but still $\delta r < \xi$ we get $F_{d=2}^{2}\propto 1/(k_{F}\delta r)$  and in one dimension there is no power-law decay as a function of $k_{F}\delta r$. In (\ref{I1}) we introduced the  Gamma function $\Gamma(x)$ and the dimensionless tunnel-conductance $\Gamma= \pi\nu_{S}\nu_{l}|t_{0}|^{2}$ with $\nu_{S}$ being the DOS per volume of the SC at the Fermi level $\mu_{S}$.
The result shows the well known power-law decay at low bias $\mu$ characteristic of dynamical CB \cite{Devoret}. The exponent $4/g$ in (\ref{I1}) is two times the power for single electron  tunneling via  one junction. This is so, because the two tunneling events are not correlated since each electron tunnels to a different lead and the charge relaxation process for each circuit is independent.\\
We consider now the large bias regime $\Delta, \mu\gg\omega_{R}$. In the regime $\Delta,|\mu-E_{c}|\gg\omega_{R}$ we can use the short time expansion for $J(t)$ in (\ref{corr1}). As long as $|\mu-E_{c}|\ll\Delta$ we can again neglect the delay times $t'$ and $t''$ compared to $t$ in all correlation functions in (\ref{corr1}) and obtain the current $I_{1}$ in the large bias limit and up to small contributions $\sim e\pi\Gamma^{2}F_{d}^{2}(\delta r)\omega_{R}[{\cal O}(\omega_{R}/\mu)+{\cal O}(\omega_{R}/|\mu-E_{c}|)]$
\begin{equation}
I_{1}=e\pi\Gamma^{2}F_{d}^{2}(\delta r)\Theta(\mu-E_{c})(\mu-E_{c}).
\end{equation}
This shows the development of a gap in $I_{1}$ for $\mu<E_{c}$ and $R\rightarrow \infty$ which is a striking feature of the dynamical CB.\\
{\it Current of two electrons tunneling into the same lead.--\/}
We turn now to the calculation of the current $I_{2}$ carried by spin-entangled electrons that tunnel both into the {\it same} lead either 1 or 2. The current formula for $I_{2}$ is given by (\ref{current1}) but with $n=n'$ and $m=m'=n$, and we assume that the two electrons tunnel off the SC from the same point and therefore $\delta r=0$ here. Since both electrons tunnel into the same lead the bath correlation functions do not separate anymore as was the case in (\ref{corr1}). Instead we have to look at the full 4-point correlator
\begin{eqnarray}
\label{bathcorr}
\langle e^{i{\phi}_{n}(t-t'')} e^{i{\phi}_{n}(t)} e^{-i{\phi}_{n}(t')} e^{-i{\phi}_{n}(0)}\rangle&&\nonumber\\
=\frac{e^{J(t-t'-t'')+J(t-t')+J(t-t'')+J(t)}}{e^{J(t')+J(-t'')}}.&&
\end{eqnarray}
The lead correlators again factorize into a product of two single-particle Green's functions since they are assumed to be Fermi liquids and in addition there appear  no spin correlations due to tunneling of two electrons with opposite spins. \\
We first consider the low bias regime $\mu\ll\omega_{R},\Delta$. Here again we can assume that $|t|$ is large compared to the delay times $t'$ and $t''$, but it turns out to be crucial to distinguish carefully between $\Delta\gg\omega_{R}$ and $\Delta\ll\omega_{R}$.
We first treat the case $\Delta\gg \omega_{R}$ and approximate 
$\exp[-(J(t')+J(-t''))]\simeq\exp[-iE_{c}(t''-t')]$ in (\ref{bathcorr}). In this limit and for $\Delta>E_{c}$ the current $I_{2}$ becomes
\begin{eqnarray}
\label{largeD}
I_{2}&=&e{\pi}\mu\Gamma^{2}\frac{(4\Delta/\pi)^{2}}{\Delta^{2}-E_{c}^{2}}\arctan^{2}\sqrt{\frac{\Delta+E_{c}}{\Delta-E_{c}}}\nonumber\\
&&\times\frac{e^{-8\gamma/g}}{\Gamma(8/g+2)}\left(\frac{2\mu}{\omega_{R}}\right)^{8/g}.
\end{eqnarray}
The exponent  $8/g$ in (\ref{largeD}) we would also get in a {\it first-order} tunneling event if the operator $e^{-i{\phi}_{n}}$  is replaced by $e^{-i2{\phi}_{n}}$ in (\ref{Htunnel}) which changes the charge of the tunnel junction capacitor $n$ by 2e. In addition to this {\it double} charging effect we find from (\ref{largeD}) that  an enhancement of $E_{c}$ gives not only rise to a suppression of $I_{2}$ via the term 
$(2\mu/\omega_{R})^{8/g}=(2\mu\pi/gE_{c})^{8/g}$ but also to an increase due to the $\Delta$-dependent prefactor. 
This enhancement can be interpreted as a relaxation of the charge imbalance created by the first electron tunneling event at small times, much smaller than the classical relaxation time $\tau_{cl}$. 
The result (\ref{largeD}) is valid  if $\sqrt{(\Delta-E_{c})/(\Delta+E_{c})}\gg\sqrt{\omega_{R}/\Delta}$.\\
In the other limit where $\Delta\ll\omega_{R}$, e.g. for small $R$,  we can assume that $\omega_{R}t'$ and $\omega_{R}t''\gg 1$ and therefore we approximate $\exp[-(J(t')+J(-t''))]\simeq \exp(-4\gamma/g)\omega_{R}^{4/g}(t't'')^{2/g}$. In this limit we obtain
\begin{equation}
\label{smallD}
I_{2}=e\pi\mu\Gamma^{2}A(g)\left(\frac{2\mu}{\omega_{R}}\right)^{4/g}\left(\frac{2\mu}{\Delta}\right)^{4/g}, 
\end{equation}
with $A(g)=(2e^{-\gamma})^{4/g}\Gamma^{4}(1/g+1/2)/\pi^{2}\Gamma(8/g+2)$.
Here the relative suppression of the current $I_{2}$ compared to $I_{1}$ is given essentially by $(2\mu/\Delta)^{4/g}$ and not by $(2\mu/\omega_{R})^{4/g}$ as in the case of an {\it infinite} $\Delta$. This is because the virtual state with a quasiparticle in the SC can last much longer than the classical relaxation time $\tau_{cl}$, and, as a consequence,  the power law suppression of the current is weakened since $\Delta\ll\omega_{R}$ here. To our knowledge, the result (\ref{smallD}) was not discussed in the literature so far \cite{comment}, but similar results are obtained when SCs are coupled to Luttinger liquids \cite{RL}. It is important to note that a large gap $\Delta$ is therefore crucial to suppress $I_{2}$.\\
In the large  voltage regime $\Delta,\mu\gg\omega_{R}$ we expect a Coulomb gap due to a charge $q=2e$. Indeed, in the parameter range $|\mu-2E_{c}|\gg\omega_{R}$ and $\Delta\gg|\mu-E_{c}|$ we obtain $I_{2}$ again up to small contributions $\sim e\pi\Gamma^{2}\omega_{R}[{\cal O}(\omega_{R}/\mu)+{\cal O}(\omega_{R}/|\mu-2E_{c}|)]$
\begin{equation}
I_{2}=e\pi\Gamma^{2}\Theta(\mu-2E_{c})(\mu-2E_{c}).
\end{equation}
This shows that $I_{2}$ is small ($\propto \omega_{R}^{2}/|\mu-2E_{c}|$) in the regime $E_{c}<\mu<2E_{c}$, whereas $I_{1}$ is finite ($\propto F_{d}^{2}(\delta r)(\mu-E_{c})$).\\
{\it Discussion and conclusions.--\/}
We now give numerical values for the current magnitudes and efficiencies of our entangler.
We first discuss the low bias regime $\mu\ll\Delta,\omega_{R}$. In Fig. 2 we show the ratio $I_{2}/I_{1}$ (efficiency of entangler) and  $I_{1}$  for $\Delta\gg E_{c},\omega_{R}$ as a function of $4/g$ for realistic system parameters (see figure caption).
The plots show that a very efficient entangler can be expected for lead resistances on the order of $R\stackrel{<}{\sim}R_{Q}$. The total current is then on the order of $I_{1}\stackrel{>}{\sim}10$ fA.
In the large bias regime $\mu\gg\omega_{R}$ and for $E_{c}<\mu< 2E_{c}$ we obtain $I_{2}/I_{1}\propto (k_{F}\delta r)^{d-1}\omega_{R}^{2}/(2E_{c}-\mu)(\mu-E_{c})$, where we assume that $2E_{c}-\mu$ and $\mu-E_{c}\gg\omega_{R}$. For $\mu\simeq 1.5E_{c}$ and using $\omega_{R}=gE_{c}/\pi$ we obtain approximately $I_{2}/I_{1}\propto (k_{F}\delta r)^{d-1}g^{2}$. 
\begin{figure}[h]
\vspace{0.3cm}
\begin{center}
\hbox{\resizebox{4.5cm}{!}{\includegraphics{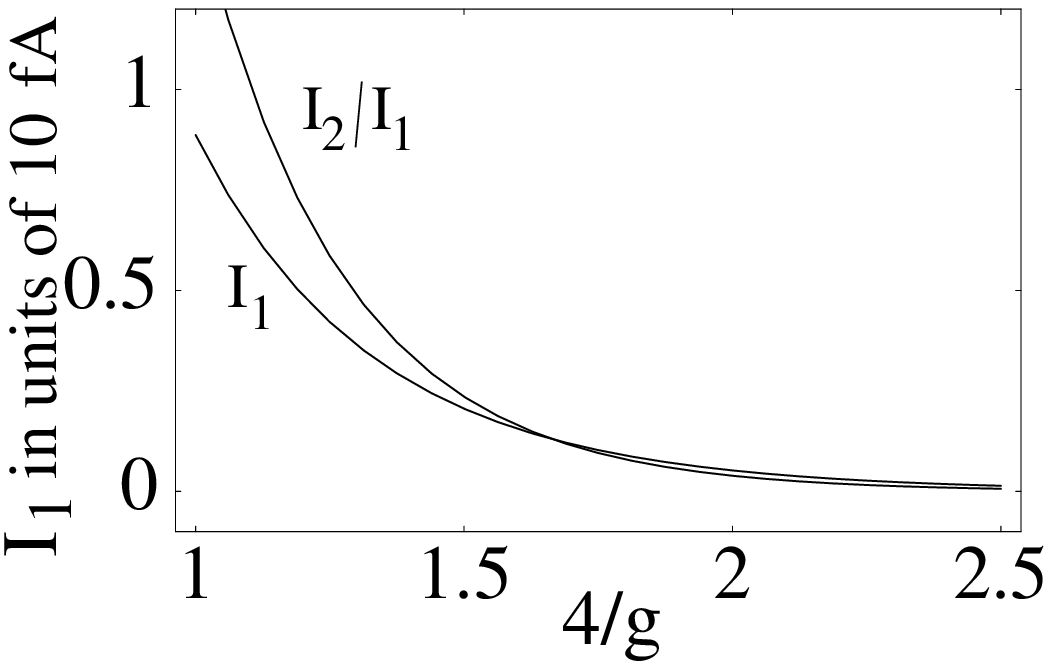}}\,\resizebox{4.2cm}{!}{\includegraphics{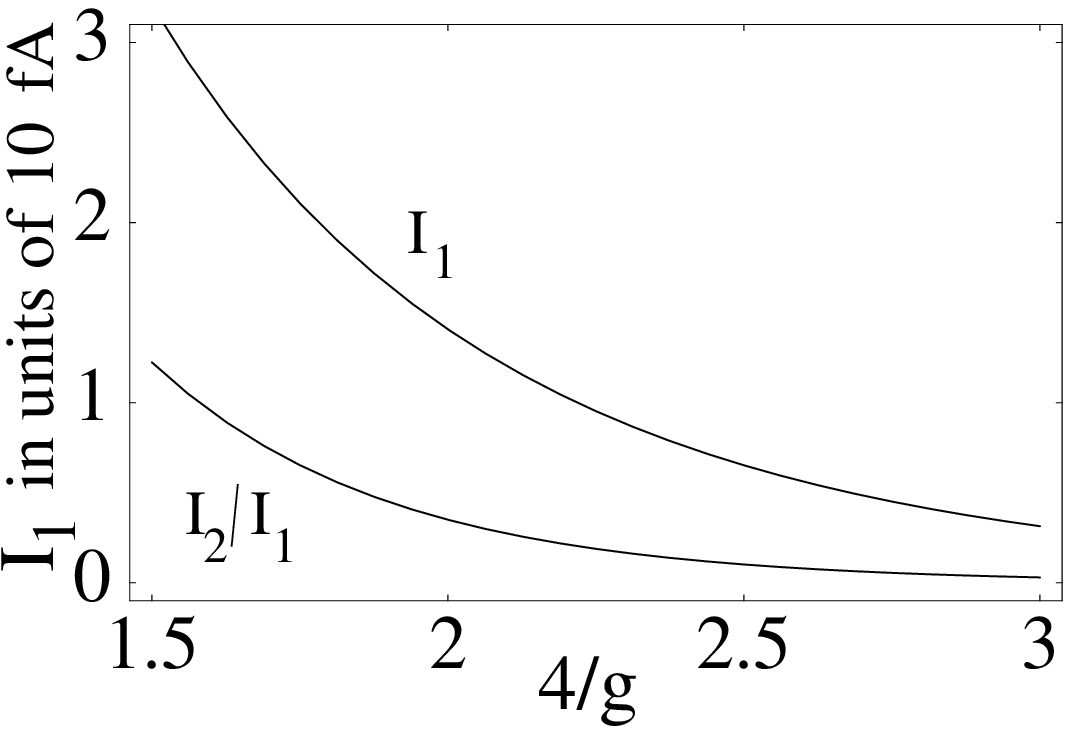}}}
\end{center}
\vspace{-0.3cm}
\caption{Current ratio $I_{2}/I_{1}$ (entangler efficiency) and current $I_{1}$ in the low bias regime, $\mu\ll\Delta,\omega_{R}$ and $\Delta\gg E_{c},\omega_{R}$, as a function of $4/g=4R/R_{Q}$. We have chosen realistic parameters: $E_{c}=0.1$ meV, $k_{F}\delta r=10$, $\Gamma=0.1$. The left plot is for $\mu=5$ $\mu$eV and the right one for $\mu=15$ $\mu$eV. In the case of a 2d SC, $I_{1}$ and $I_{1}/I_{2}$ can be multiplied by 10.}
\end{figure}
To have $I_{2}/I_{1}<1$ we have to achieve $g^{2}<0.01$ for $d=3$, and $g^{2}<0.1$ for $d=2$. Such small values of $g$ have been produced approximately in Cr leads \cite{Kuzmin}. For  $I_{1}$ we obtain $I_{1}\simeq e(k_{F}\delta r)^{1-d}(\mu-E_{c})\Gamma^{2}\simeq e(k_{F}\delta r)^{1-d}E_{c}\Gamma^{2}\simeq 2.5$ pA for $d=3$ and for the same parameters as  used in Fig. 2. 
This shows that here $I_{1}$ is much larger than for low bias voltages, but to have an efficient entangler very high lead resistances on the order $R \stackrel{>}{\sim}10R_{Q}$ should be used.
Our discussion shows that it should be possible to implement the proposed device within state of the art techniques.\\
{\it Acknowledgements.} 
We acknowledge useful discussions with C. Bruder, H. Gassmann, and F. Marquardt. This work was supported by the Swiss NSF, NCCR Basel, DARPA, and ARO.


\begin{references}

\bibitem{Chuang}
M. Nielsen, I. Chuang, {\it Quantum Computation and Quantum Information}, Cambridge University Press, 2000.

\bibitem{Bell}
J.S. Bell, Rev. Mod. Phys. {\bf 38}, 447 (1966).

\bibitem{RSL}
P. Recher, E.V. Sukhorukov, D. Loss, Phys.\ Rev.\ B {\bf 63},
165314 (2001).

\bibitem{RL}
P. Recher, D. Loss, Phys. Rev. B {\bf 65}, 165327 (2002).

\bibitem{Yamamoto}
W.D. Oliver, F. Yamaguchi, Y. Yamamoto, Phys. Rev. Lett. {\bf 88}, 037901 (2002).

\bibitem{Saraga}
D.S. Saraga, D. Loss, Phys. Rev. Lett. {\bf 90}, 166803 (2003).

\bibitem{Bena}
C. Bena {\it et al.}, Phys. Rev. Lett. {\bf 89}, 037901 (2002).

\bibitem{tanglers}
G.B. Lesovik, T. Martin, G. Blatter, Eur. Phys. J. B {\bf 24}, 287 (2001); S. Bose, D. Home, Phys. Rev. Lett. {\bf 88}, 050401 (2002); P. Samuelsson, E.V. Sukhorukov, M. B\"uttiker, cond-mat/0303531; C.W.J. Beenakker {\it et al.}, cond-mat/0305110.

\bibitem{Loss}
D. Loss, D.P. DiVincenzo, Phys. Rev. A {\bf 57}, 120
(1998).

\bibitem{Choi}
M.-S. Choi, C. Bruder, D. Loss, Phys. Rev. B {\bf 62}, 13569, (2000).

\bibitem{Awschalom}  
J.M. Kikkawa, D.D. Awschalom, Phys. Rev. Lett. {\bf 80}, 4313 (1998).

\bibitem{InAs}
J. Nitta {\it et al.}, Phys. Rev. B {\bf 46}, 1486 (1992); C. Nguyen, H. Kroemer, E.L. Hu, Phys. Rev. Lett. {\bf 69}, 2847 (1992).

\bibitem{Franceschi}
S. De Franceschi {\it et al.}, Appl. Phys. Lett. {\bf 73}, 3890 (1998).

\bibitem{Marsh}
A.M. Marsh, D.A. Williams, H. Ahmed, Phys. Rev. B {\bf 50}, 8118 (1994).


\bibitem{Rimberg}
A.J. Rimberg {\it et al.}, Phys. Rev. Lett. {\bf 78}, 2632 (1997).

\bibitem{Kuzmin}
L.S. Kuzmin {\it et al.}, Phys. Rev. Lett. {\bf 67}, 1161 (1991).

\bibitem{Diener}
J.J. Hesse, G. Diener, Physica B {\bf 203}, 393 (1994).

\bibitem{Huck}
A. Huck, F.W.J. Hekking, B. Kramer, Eur. Phys. Lett. {\bf 41}, 201 (1998).

\bibitem{splitter}
The degree of entanglement could be tested in terms of noise in a beam splitter setup \cite{noise}.


\bibitem{noise}
G. Burkard, D. Loss, E.V. Sukhorukov, Phys. Rev. B {\bf 61}, R16303 (2000).

\bibitem{leads}
To prevent a current flowing from one lead to the other via the SC.

\bibitem{Devoret}
M.H. Devoret {\it et al.}, Phys. Rev. Lett. {\bf 64}, 1824 (1990); see also G.-L. Ingold and Y.V. Nazarov, ch. 2 in H. Grabert and M.H. Devoret (eds.), Single Charge Tunneling, Plenum Press, New York, 1992.



\bibitem{Leggett}
A.O. Caldeira, A.J. Leggett, Ann. Phys. (N.Y.) {\bf 149}, 374 (1983).

\bibitem{oscillators}
Any lead impedance $Z_{n}(\omega)$  can be modeled with Eq. (\ref{env}) via $Z^{-1}_{n}=\int_{-\infty}^{+\infty}dt\exp(-i\omega t)Y_{n}(t)$ where the admittance $Y_{n}(t)=\sum_{j=1}^{N}(\Theta(t)/L_{nj})\cos(t/\sqrt{L_{nj}C_{nj}})$.

\bibitem{comment1}
A correction to this decoupling assumption is determined by the cross-capacitance $C_{12}\ll C_{n}$, $n=1,2$ between the leads 1,2 and is therefore small.

\bibitem{Merzbacher}
E. Merzbacher, {\it Quantum Mechanics} 3rd ed., John
Wiley and Sons, New York, 1998, ch. 20.


\bibitem{dimension}
It was predicted \cite{Klapwijk} and shown experimentally \cite{Weiss} that a SC on top of a 2DEG can induce superconductivity in the 2DEG via the proximity effect leading to an effectively two dimensional SC.

\bibitem{Klapwijk}
A.F. Volkov {\it et al.}, Physica \ C {\bf 242}, 261 (1995).

\bibitem{Weiss}
J. Eroms {\it et al.}, Eur. Phys. Lett. {\bf 58}, 569 (2002).

\bibitem{comment}
The result (\ref{smallD}) is in contrast to predictions  made in \cite{Diener}.

\end{references}
\end{document}